\documentclass{article}
\usepackage{spconf,amsmath,graphicx,booktabs,array}
\usepackage[font=normalsize,labelfont=bf]{caption}
\usepackage{floatrow} 
\title{Automating Vitiligo Skin Lesion Segmentation Using Convolutional Neural Networks}
\name{Makena Low, Priyanka Raina}
\address{Stanford University}
\usepackage{xcolor}

\iftrue  % all comments
    \newcommand{\COMMENT}[3]{{\color{#1}{[}{~#2:~#3~}{]}}}
\else
    \hypersetup{draft}
    \newcommand{\COMMENT}[3]{}
\fi

\begin{document}
%\ninept
%
\maketitle
%

%\jz{hello}

\begin{abstract}
For several skin conditions such as vitiligo, accurate segmentation of lesions from skin images is the primary measure of disease progression and severity. Existing methods for vitiligo lesion segmentation require manual intervention. Unfortunately, manual segmentation is time and labor-intensive, as well as irreproducible between physicians. We introduce a convolutional neural network (CNN) that quickly and robustly performs vitiligo skin lesion segmentation. Our CNN has a U-Net architecture with a modified contracting path. We use the CNN to generate an initial segmentation of the lesion, then refine it by running the watershed algorithm on high-confidence pixels. We train the network on 247 images with a variety of lesion sizes, complexity, and anatomical sites. The network with our modifications noticeably outperforms the state-of-the-art U-Net, with a Jaccard Index (JI) score of 73.6\% (compared to 36.7\%). Moreover, our method requires only a few seconds for segmentation, in contrast with the previously proposed semi-autonomous watershed approach, which requires 2-29 minutes per image.

\end{abstract}
\section{Introduction}
\begin{figure}[t]
    \centering
    \includegraphics[width=1.0 \linewidth]{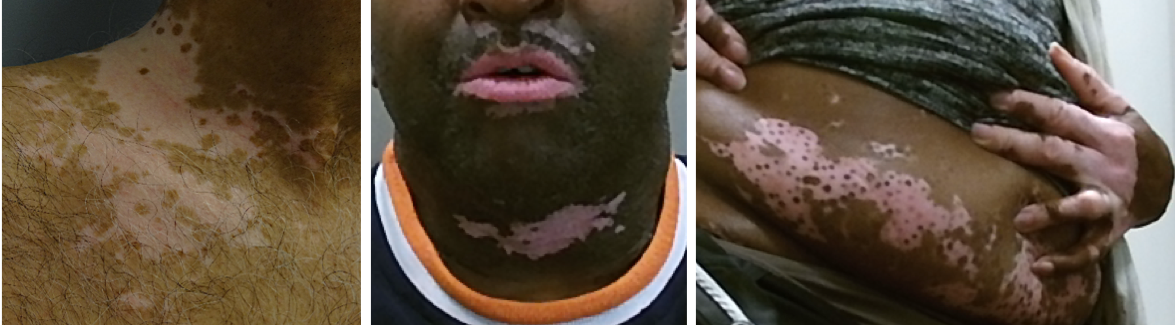}
    \caption{Examples of vitiligo lesions with different sizes, complexity, and anatomical sites.}
    \label{fig:tree}
\end{figure}

Vitiligo is a skin condition where patches of skin get depigmented, as shown in Fig.~\ref{fig:tree}. It affects 0.5-2\% of the population, can be developed by anyone, and though not physically painful, can harm patients psychologically, socially, and professionally \cite{hazel-jemmott_hazel-jemmott_2016}\cite{amer2016quality}\cite{salzes2016vitiligo}. The body surface area (BSA) affected by vitiligo is the main measure of the condition's severity and progression. BSA measurements must be consistent for proper clinical care, translational research efforts, and assessment of the efficacy of treatment. For instance, a physician's visual estimation of the percentage of vitiligo-affected BSA informs the Vitiligo Area Scoring Index (VASI) and Vitiligo European Task Force (VETF) metrics. Both measures can only detect large changes in lesion area, with the smallest being between 7.1\% to 10.4\% of total BSA \cite{komen2015vitiligo}.
Current segmentation practices are mainly manual. Not only is this method detrimental for accurate and reproducible readings, but it is also a time-inefficient and labor-intensive process. Moreover, non-dermatologists are often the ones who perform these segmentations, even though they do not have a rigorous background for such reviews \cite{raina2018energy}. This study aims to introduce a novel solution to this issue. 

A convolutional neural network (CNN) is a promising approach for solving complex skin segmentation challenges. CNNs for skin cancer segmentation are already in widespread use, in large part due to the International Skin Imaging Collaboration (ISIC) Skin Lesion Analysis Towards Melanoma Detection competition \cite{codella2018skin}. However, vitiligo is seldom the subject of such segmentation studies. One study that uses CNNs for vitiligo segmentation is very data-intensive: it presents a model trained on about 40,000 images, which is much larger than our and most medical datasets \cite{liu2019classification}. 

\begin{figure}[b]
    \centering
    \includegraphics[width=1.0\linewidth]{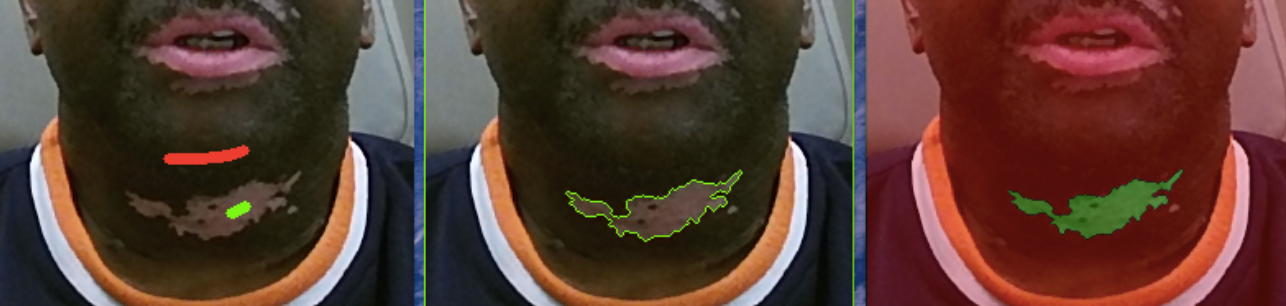}
    \caption{Illustration of watershed algorithm with manual seeding (left), the resulting contour (middle), and segmented output (right).}
    \label{fig:watershed}
\end{figure}

Researchers have also explored less computationally intensive techniques than CNNs. One study attempted to quantify treatment efficacy by using a computerized digital imaging analysis system (C‐DIAS) \cite{shamsudin2015objective}. Sheth et al. leveraged standard color image processing techniques to create an automatic vitiligo segmentation program; however, this approach does not perform well when tested on large surface areas \cite{sheth2015pilot}. To address this, Raina et al. created a graphical user interface (GUI) with a semi-autonomous version of the watershed algorithm for lesion segmentation \cite{raina2018energy}\cite{roerdink2000watershed}. The tool succeeds in outputting subtle contours for full-body images, but it requires ``seeds" from the user to define the background (environment and healthy skin) and foreground (affected skin), as shown in Fig.~\ref{fig:watershed} (left) in red and green colors. This semi-manual process of segmentation requires significant work when lesions are involved, as shown in Fig.~\ref{fig:tree} (right). Our work addresses these shortcomings.

We introduce a CNN that achieves a high Jaccard Index score (intersection over union) of 73.6\% with 247 training images. Our models are based on the end-to-end U-Net \cite{ronneberger2015u} architecture. We substitute the contracting path with a popular semantic segmentation CNN that serves as a feature extractor. Our work investigates VGG16, ResNet50, InceptionV3, InceptionResNetV2, and SENet154 as contracting path enhancers \cite{simonyan2014very}\cite{he2016deep}\cite{szegedy2016rethinking}\cite{szegedy2017inception}\cite{hu2018squeeze}. We also experiment with watershed-based post-processing; after classification, the high-confidence pixels are fed as seeds to the watershed algorithm \cite{roerdink2000watershed}. We find that an InceptionResnetV2 contracting path performs the best out of all our explored architectures. Our method drastically reduces segmentation time compared to the watershed GUI as well as offers a method of achieving reproducible output.

\section{Methods}

\subsection{Vitiligo Image Samples and Annotation}
Our dataset consists of 308 red/green/blue (RGB) images of vitiligo lesions compiled by the UC Davis Medical Center. The lesions range widely in skin tone and anatomical location. Physicians have taken the images from several angles, at different levels of brightness, and either in ultraviolet (UV) or natural lighting. We derive the ground truth segmentation output from the semi-autonomous watershed GUI and manual edits. Each ground truth output image is a binary mask of the lesion, where zero (black) represents healthy skin or the environment, and 255 (white) represents vitiligo. The dataset is split such that 60\% is for training the model (188 images), 20\% is for validating the model (66 images), and 20\% for testing the model on unseen data (61 images). 

\begin{figure}[t]
  \includegraphics[width=\textwidth]{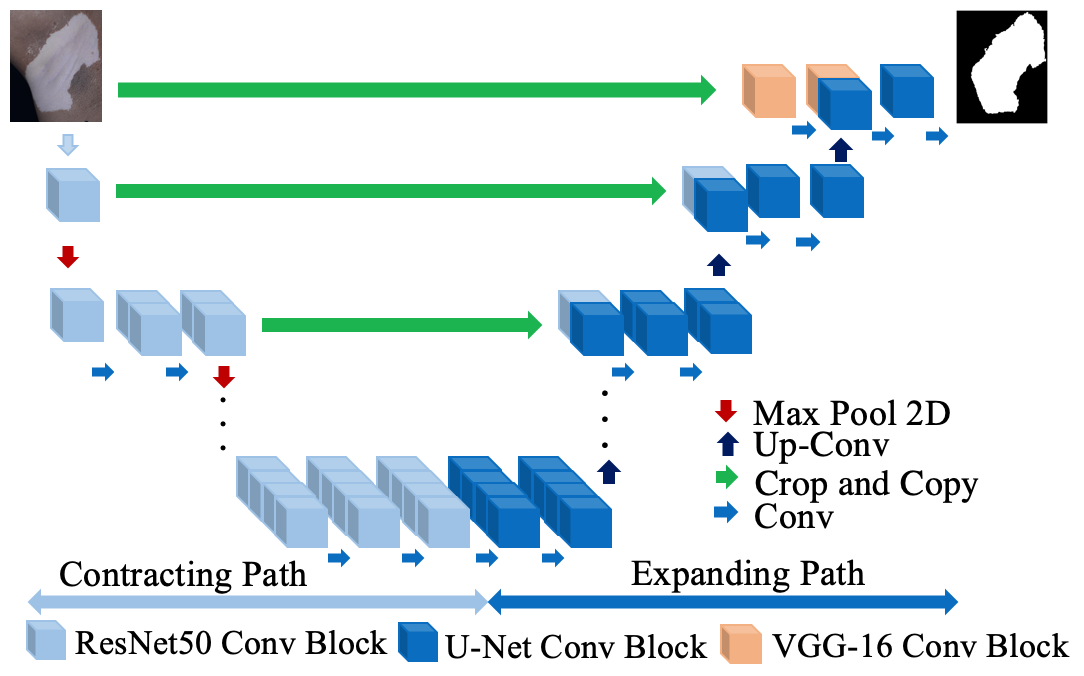}
    
  \caption{U-Net with a ResNet50 contracting path.}
  \label{fig:resnetUnet}
\end{figure}
\vspace{-0.3cm}
\subsection{Evaluation Metric}
We evaluate the network's performance using the pixel-wise Intersection over Union metric (IoU):
$$IoU = \frac{A \cap B}{A \cup B} $$
This metric is also known as the Jaccard Index (JI) \cite{codella2018skin}. For each image, we calculate the JI between every classified pixel and the corresponding ground truth pixel. The JI of the image is the average of the pixel-wise JI scores. Previous analysis suggests that the JI is too optimistic by not accounting for the labor required to correct an inaccurate segmentation \cite{codella2017deep}. Thus, we also compute a thresholded Jaccard Index to account for segmentations that do not fall within professional inter-observer variability. If the average JI is less than 65\%, we set the score to 0\% for the image. Otherwise, the JI is unchanged. The threshold of 65\% was determined by ISIC \cite{codella2018skin}. Although ISIC focuses on melanoma segmentation, the human labor required for a similar evaluation with vitiligo was not feasible for us; we suppose that the ISIC threshold is a fair estimate. The evaluation metric for our networks is the average of the threshold JI scores for the images in the validation set.
\vspace{-0.3cm}
\subsection{Image Pre-processing}
We perform simple pre-processing on images before feeding them into our network. We subtract the mean from each image channel and normalize each channel to make the standard deviation -1 to 1 to guarantee pixel scale standardization. We re-scale every image to $224\times 224$. We implement data augmentation during training (after pre-processing). Data augmentation includes a rotation range from 0 to 180 degrees, horizontal and vertical shifts set to 0.05, and vertical and horizontal flips. Moreover, we set the zoom range to 0.8 to 1.2 times the original image due to the varying closeness between camera and lesion. Due to the varying brightness and lighting conditions, brightness augmentation ranges from 0.7 to 1.3 times that of the original image. 
\vspace{-0.3cm}
\subsection{U-Net Network Experiments}
Our baseline is an unmodified U-Net with 512 hidden units at the bottleneck and no pre-trained weights from ImageNet \cite{ronneberger2015u}. The final activation is a softmax layer. After 100 epochs, the JI score is 36.7\%. We experiment with using popular semantic segmentation networks such as VGG16 and ResNet50 as modified contracting paths in our U-Net. Fig.~\ref{fig:resnetUnet} illustrates our U-Net architecture with a ResNet50 contracting path. We utilize an API based on Keras and Tensorflow frameworks to create our test architectures listed in Table~\ref{table:JI_Unets}. For fast comparison, each modified U-Net is only trained for 30 epochs and evaluated. Table~\ref{table:JI_Unets} shows the results of each model.

\begin{table}[t]
\begin{tabular}{|l|l|l|l|}
\hline
\textbf{Contracting Path}  & \textbf{Epochs} & \textbf{Val} & \textbf{Train} \\ \hline
Unmodified        & 100             & 36.8\%       & 44.7\%         \\ \hline
VGG16             & 30              & 61.2\%       & 63.7\%         \\ \hline
ResNet50          & 30              & 64.2\%       & 68.2\%         \\ \hline
InceptionV3       & 30              & 61.5\%       & 63.9\%         \\ \hline
InceptionResNetV2 & 30              & 70.9\%       & 67.0\%         \\ \hline
SENet154          & 30              & 61.3\%       & 66.7\%         \\ \hline
\end{tabular}
\caption{JI scores of U-Net architectures.}
\label{table:JI_Unets}
\end{table}
\vspace{-0.3cm}
\subsection{Hyperparameter Tuning}
Since there are benefits to multiple methods of hyperparameter tuning, we use a three-pronged approach for finding optimal hyperparameters. (1) For initial exploration, we iterate with random search to leverage its strength in not fixating on local minima while also efficiently exploring the hyperparameter search space \cite{bergstra2012random}. (2) Once coarse parameter tuning identifies promising ranges, we manually alter our search space for fine-tuning. (3) Finally, we employ sequential model-based optimization (SMBO), so we can try future hyperparameters based on promising past ones, as well as reduce the computational expense and iterations needed for promising results compared to random search \cite{snoek2012practical}. Table ~\ref{table:hp_choices} outlines our chosen hyperparameters from this optimization.

\begin{figure}[t]
    \centering
    \includegraphics[width=0.8\linewidth]{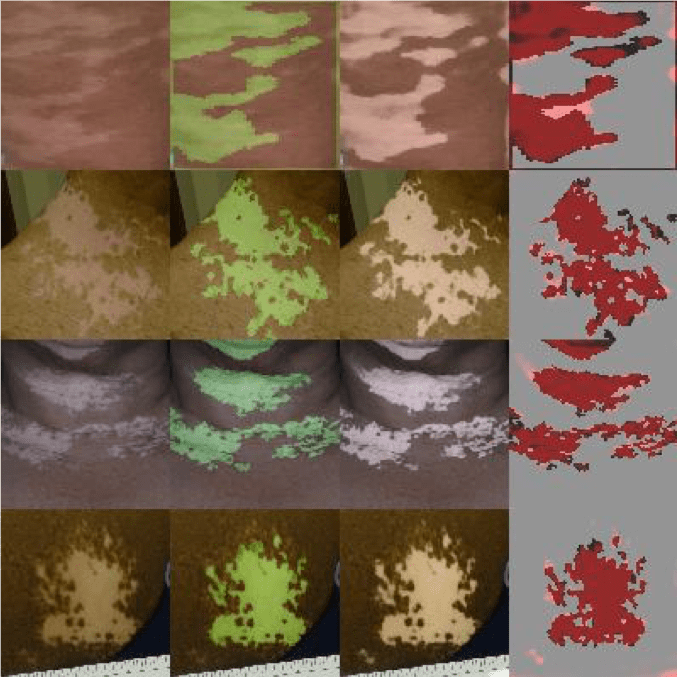}
    \caption{Original image (left), ground truth overlay (middle left), prediction overlay (middle right), ground truth overlay with prediction (red is true positive and pink is false positive) (right).}
    \label{fig:prediction-results}
\end{figure}

\begin{table}[t]
\begin{tabular}{|l|l|}
\hline
\textbf{Hyperparameter} & \textbf{Value}          \\ \hline
Loss                    & BCE-JI                  \\ \hline
LR                      & 0.000336375             \\ \hline
Optimizer               & Nadam                   \\ \hline
Contracting Normalization   & Batch                   \\ \hline
Contracting Hidden Units    & {[}512,256,128,64,32{]} \\ \hline
Freeze Weights          & False                   \\ \hline
Contracting Activation      & ELU                     \\ \hline
Weight Decay            & 0.000158                \\ \hline
Dropout                 & 0.0136                  \\ \hline
LR Decay                & 8.806E-05               \\ \hline
Epochs                  & 165                     \\ \hline
Batch Size              & 8                       \\ \hline
\end{tabular}
\caption{Tuned hyperparameters for U-Net with InceptionResNetV2 contracting path.}
\label{table:hp_choices}
\end{table}
\vspace{-0.3cm}
\subsection{Combining Datasets and Post-Processing}
We combine the training and validation sets - for a total of 247 images - to train our network before evaluating it on the test set. We also experiment with watershed-based post-processing, which feeds high-confidence classifications as seeds into the watershed algorithm. High confidence pixels are pixels classified within a 30\% confidence interval of being negative (0-77) or positive (179-255) for vitiligo.

\section{Results and Discussion}
InceptionResNetV2 is the best performing contracting path, as it achieves a JI of 74.1\% and threshold JI of 58.0\% before hyperparameter tuning. The runtime is 97 minutes for 100 epochs on a single NVIDIA Tesla K80 GPU. SENet154 also appears to be a strong candidate; however, because the high performance came at the expense of increased training time, we did not explore it further in our study. After hyperparameter tuning, the JI score is 81.5\%, and the threshold JI is 62.8\%. Once we perform watershed post-processing, the count of images below the threshold falls from 16 images to 14 images. After training on the combined dataset, the InceptionResNetV2-based U-Net achieves a JI of 73.6\% and threshold JI of 61.9\%. Fig.~\ref{fig:prediction-results} shows an example of the output. Though it is counterintuitive that performance decreases with our larger dataset, we believe this result may be due to the variability inherent in our small test set, which is only 61 images. The total training runtime is 108 minutes for about 200 epochs for our final network.

\newcommand{\easypic}{\includegraphics[width=5em,height=5em]{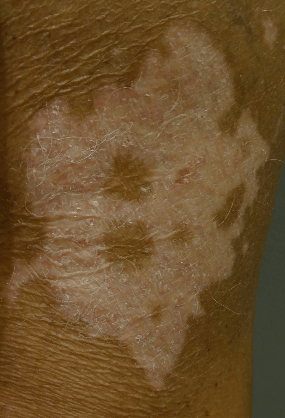}}
\newcommand{\easymaskOne}{\includegraphics[width=5em,height=5em]{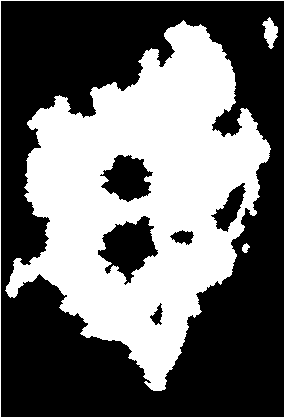}}
\newcommand{\easyMaskTwo}{\includegraphics[width=5em,height=5em]{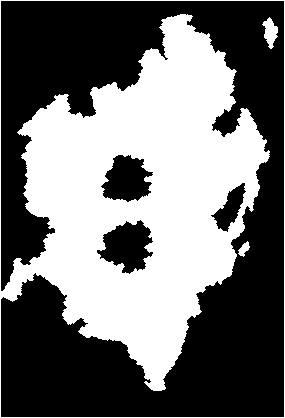}}
\newcommand{\easyMaskThree}{\includegraphics[width=5em,height=5em]{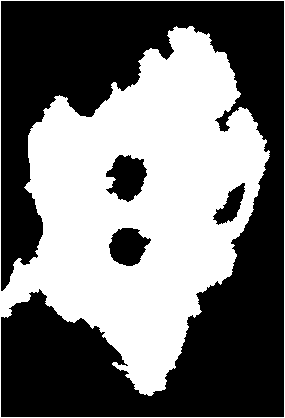}}

\newcommand{\medpic}{\includegraphics[width=5em,height=5em]{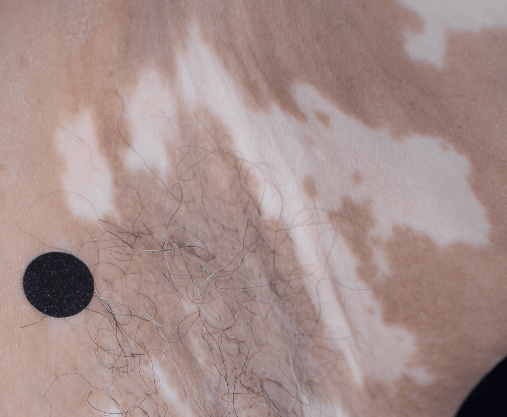}}
\newcommand{\medMaskOne}{\includegraphics[width=5em,height=5em]{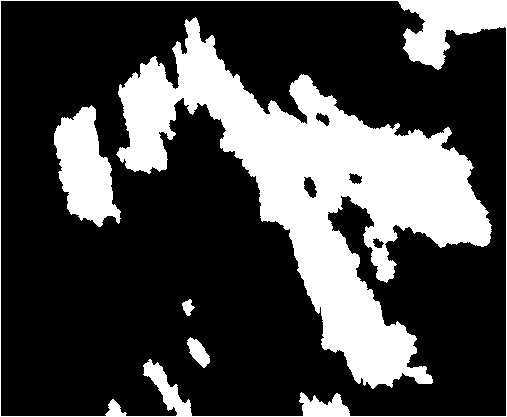}}
\newcommand{\medMaskTwo}{\includegraphics[width=5em,height=5em]{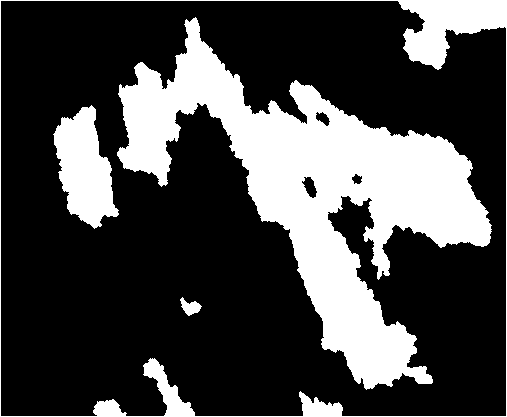}}
\newcommand{\medMaskThree}{\includegraphics[width=5em,height=5em]{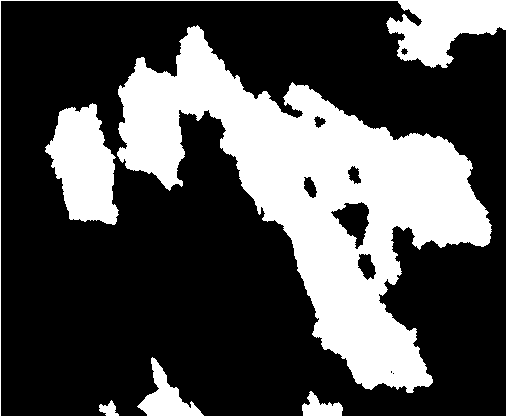}}

\newcommand{\compic}{\includegraphics[width=5em,height=5em]{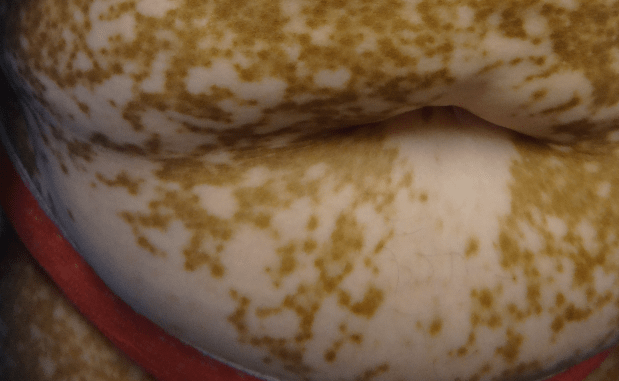}}
\newcommand{\comMaskOne}{\includegraphics[width=5em,height=5em]{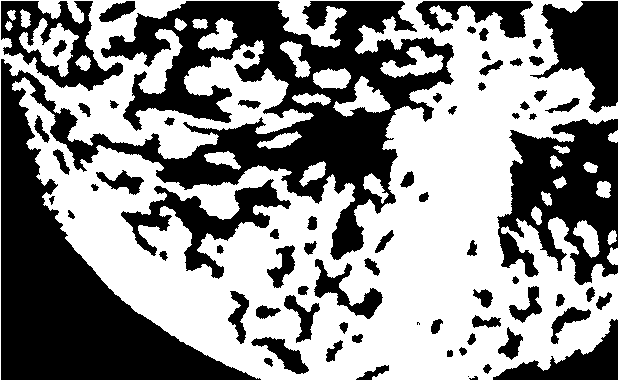}}
\newcommand{\comMaskTwo}{\includegraphics[width=5em,height=5em]{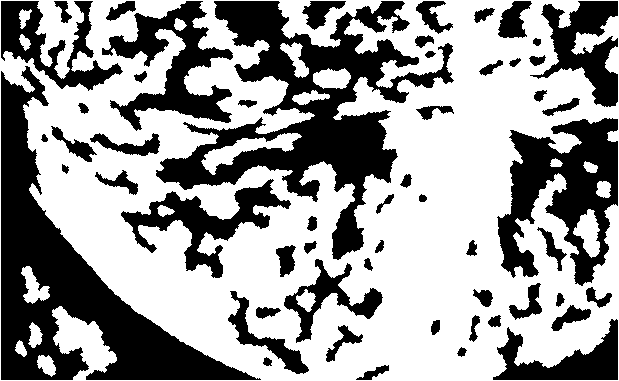}}
\newcommand{\comMaskThree}{\includegraphics[width=5em,height=5em]{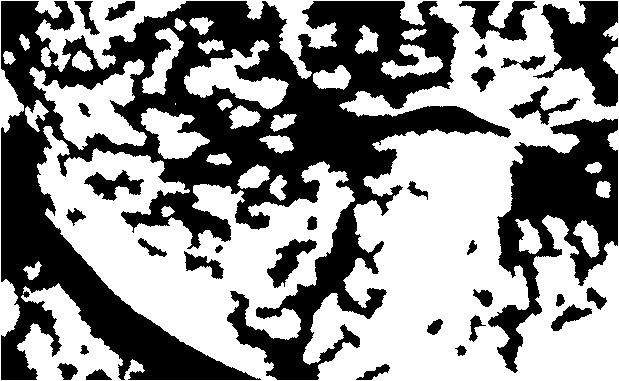}}

\newcommand{\modelMaskOne}{\includegraphics[width=5em,height=5em]{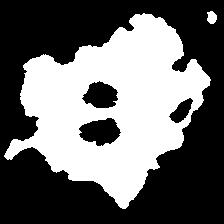}}
\newcommand{\modelMaskTwo}{\includegraphics[width=5em,height=5em]{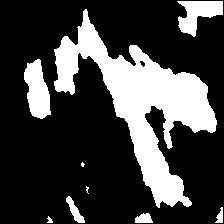}}
\newcommand{\modelMaskThree}{\includegraphics[width=5em,height=5em]{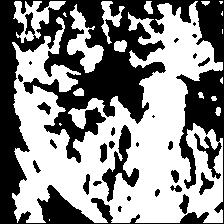}}

\newcolumntype{C}{>{\centering\arraybackslash}m{5em}}
\begin{table}[t]\sffamily
\begin{tabular}{l*4{C}@{}}
\toprule
Lesion & Simple & Moderate & Complex \\ 
\midrule
Original & \easypic & \medpic & \compic  \\ 
Our Method    &  \modelMaskOne & \modelMaskTwo & \modelMaskThree \\
Person 1 & \easymaskOne & \medMaskOne & \comMaskOne  \\ 
Person 2 & \easyMaskTwo & \medMaskTwo & \comMaskTwo  \\ 
Person 3 & \easyMaskThree & \medMaskThree & \comMaskThree  \\ 
\bottomrule 
\end{tabular}
\caption{Segmentation by our method and three persons using semi-autonomous watershed GUI  compared to the original image.}
\label{Table:segmentation-complexity-casestudy}
\end{table} 

\begin{table}[t!]
\begin{tabular}{l|l|l|l|l|}
\cline{2-5}
          & \textbf{Lesion} & \textbf{Simple} & \textbf{Moderate} & \textbf{Complex} \\ \cline{2-5} 
Our Method     & JI (\%)   & 88.7\%          & 86.1\%            & 74\%             \\ \cline{2-5} 
          & Time      & \textless{}1s   & \textless{}1s     & \textless{}1s    \\ \cline{2-5} 
Person 1 & JI (\%)   & 94.3\%          & 92.1\%            & 83.3\%           \\ \cline{2-5} 
          & Time      & 4m 55s          & 9m 4s             & 28m 46s          \\ \cline{2-5} 
Person 2 & JI (\%)   & 96.8\%          & 95.3\%            & 81.9\%           \\ \cline{2-5} 
          & Time      & 3m 44s          & 6m 57s            & 20m 39s          \\ \cline{2-5} 
Person 3 & JI (\%)   & 88.0\%          & 85.8\             & 75.6\%           \\ \cline{2-5} 
          & Time      & 1m 53s          & 4m 31s            & 17m 24s          \\ \cline{2-5} 
\end{tabular}
\caption {Segmentation scores (JI) and times for lesions of varying complexity for three persons using semi-autonomous watershed GUI compared with our method.}
\label{Table:segmentation-casestudy-scores}
\end{table}

\newcommand{\tenMinOne}{\includegraphics[width=5em,height=5em]{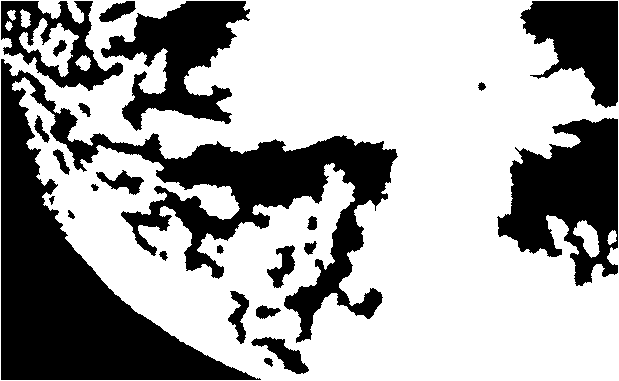}}
\newcommand{\tenMinTwo}{\includegraphics[width=5em,height=5em]{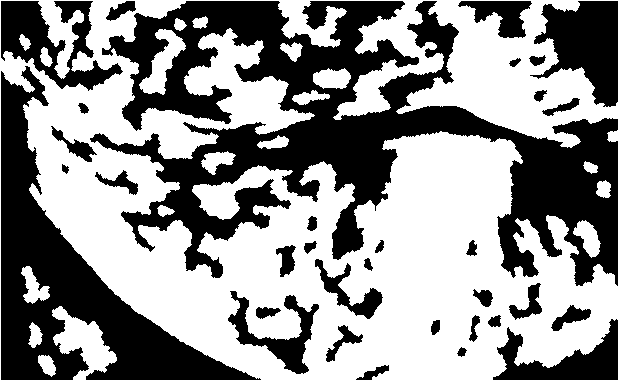}}
\newcommand{\tenMinThree}{\includegraphics[width=5em,height=5em]{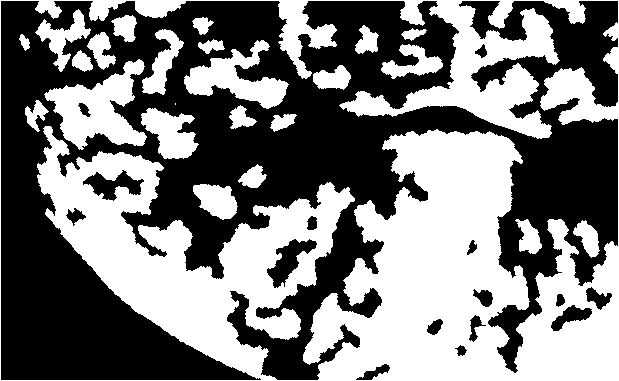}}
\newcolumntype{C}{>{\centering\arraybackslash}m{5em}}
\begin{table}[t]\sffamily
\begin{tabular}{l*4{C}@{}}
\toprule
 Lesion & Person 1 & Person 2 & Person 3 \\ 
\midrule
Complex & \tenMinOne & \tenMinTwo & \tenMinThree  \\ 
\bottomrule 
\end{tabular}
\caption{Segmentation when constrained to 10 minutes.}
\label{Table:timed-segmentation-images}
\end{table} 

\begin{table}[t]
\begin{tabular}{|l|l|l|l|}
\hline
              & \textbf{Person 1} & \textbf{Person 2} & \textbf{Person 3} \\ \hline
Accuracy (\%) & 75.9\%             & 84.7\%             &  77.6\%            \\ \hline
\end{tabular}
\caption{Variability in segmentation accuracy for the ``complex" rated lesion, with time held constant at 10 minutes.}
\label{table:timed-segmentation}
\end{table}

We conduct an error analysis on the validation images that scored below the threshold JI, 16 images in total. By inspection, we believe that eight of the images have errors primarily due to ground truth labeling limitations. The semi-autonomous watershed tool is limited in its ability to identify small lesions due to the coarseness of seeds. Manual labeling addresses some of these smaller lesions. However, there are cases in which a gradient between healthy skin to vitiligo leaves ambiguity for classification.

Moreover, pixels that are not fully confident in classification receive a lower JI score due to the way the JI is calculated. For instance, a classification of 0.7 will result in a lower JI than a classification of 1, even if both are correct in being reasonably confident that the pixel is positive for vitiligo. Still, even with errors in the labeling and a lower JI, the predictions visually capture complex target regions on any skin surface with any skin tone. The strong visual prediction suggests that the proposed architecture represents a solid foundation for future work in automating vitiligo lesion segmentation. Moreover, each segmentation took less than a few seconds per image, instead of a few minutes via semi-autonomous watershed.

We also perform a case study to quantify the correlation between lesion complexity and time to segment the lesion with the watershed GUI. We asked three persons (reviewers), who were non-dermatologists, to semi-manually segment the lesions using watershed GUI. They were allowed to gain familiarity with the GUI on practice lesions before being officially timed. We asked our reviewers to continue contouring until they felt comfortable with their segmentation being in a clinical setting. As expected, time for segmentation increased with lesion complexity. Quantitative results are shown in Table~\ref{Table:segmentation-complexity-casestudy}. Table~\ref{Table:segmentation-casestudy-scores} shows that our method requires less than a second, in contrast with watershed, which requires 2-29 minutes per image. We performed a similar case study to elucidate the variability in segmentation between reviewers. After 10 minutes of segmentation on the ``complex" rated lesion, the reviewers are asked to pause so that we can save their progress at that moment in time. From this study, we see that segmentation accuracy indeed varies widely, with almost 10\% difference between reviewers, as shown in Table~\ref{table:timed-segmentation}. Table~\ref{Table:timed-segmentation-images} visually demonstrates variability between reviewers. Our method removes this variability. 

\section{Conclusion}
We demonstrate that a U-Net with an InceptionResnetV2 - based contracting path, with watershed post-processing, proves promising for vitiligo segmentation. We quantify the variability that is possible between reviewers, as well as the time required to segment increasingly complex lesions. Our method eliminates both variability and long segmentation times, while also providing predictions that do not require much manual re-editing. There exist no conflicts of interest.

%Moreover, we can utilize the segmentations of our network to aid in the measurement of BSA. Some studies have developed a 3D analysis of vitiligo-affected BSA with VECTRA; however, these methods still require manual segmentation \cite{kohli2015three}\cite{hayashi2016novel}. We envisage our autonomous system being integrated with a method to perform 3D vitiligo lesion segmentation in future work.

%\newpage
\bibliographystyle{IEEEbib}
\bibliography{final_paper}

\begin{thebibliography}{10}

\bibitem{hazel-jemmott_hazel-jemmott_2016}
Zakiya Hazel-Jemmott and Zakiya Hazel-Jemmott,
\newblock ``Vitiligo: Causes, myths, and facts,'' Jan 2016.

\bibitem{amer2016quality}
Abdulrahman~AA Amer and Xing-Hua Gao,
\newblock ``Quality of life in patients with vitiligo: an analysis of the
  dermatology life quality index outcome over the past two decades,''
\newblock {\em International journal of dermatology}, vol. 55, no. 6, pp.
  608--614, 2016.

\bibitem{salzes2016vitiligo}
Camille Salzes, Sophie Abadie, Julien Seneschal, Maxine Whitton, Jean-Marie
  Meurant, Thomas Jouary, Fabienne Ballanger, Franck Boralevi, Alain Taieb,
  Charles Taieb, et~al.,
\newblock ``The vitiligo impact patient scale (vips): development and
  validation of a vitiligo burden assessment tool,''
\newblock {\em Journal of Investigative Dermatology}, vol. 136, no. 1, pp.
  52--58, 2016.

\bibitem{komen2015vitiligo}
L~Komen, V~Da~Graca, A~Wolkerstorfer, MA~de~Rie, CB~Terwee, and JPW van~der
  Veen,
\newblock ``Vitiligo area scoring index and vitiligo european task force
  assessment: reliable and responsive instruments to measure the degree of
  depigmentation in vitiligo,''
\newblock {\em British Journal of Dermatology}, vol. 172, no. 2, pp. 437--443,
  2015.

\bibitem{raina2018energy}
Priyanka Raina,
\newblock {\em Energy-efficient circuits and systems for computational imaging
  and vision on mobile devices},
\newblock Ph.D. thesis, Massachusetts Institute of Technology, 2018.

\bibitem{codella2018skin}
Noel~CF Codella, David Gutman, M~Emre Celebi, Brian Helba, Michael~A Marchetti,
  Stephen~W Dusza, Aadi Kalloo, Konstantinos Liopyris, Nabin Mishra, Harald
  Kittler, et~al.,
\newblock ``Skin lesion analysis toward melanoma detection: A challenge at the
  2017 international symposium on biomedical imaging (isbi), hosted by the
  international skin imaging collaboration (isic),''
\newblock in {\em 2018 IEEE 15th International Symposium on Biomedical Imaging
  (ISBI 2018)}. IEEE, 2018, pp. 168--172.

\bibitem{liu2019classification}
Jian Liu, Jianwei Yan, Jie Chen, Guozhong Sun, and Wei Luo,
\newblock ``Classification of vitiligo based on convolutional neural network,''
\newblock in {\em International Conference on Artificial Intelligence and
  Security}. Springer, 2019, pp. 214--223.

\bibitem{shamsudin2015objective}
Norashikin Shamsudin, Suraiya~H Hussein, Hermawan Nugroho, and Mohd~Hani
  Ahmad~Fadzil,
\newblock ``Objective assessment of vitiligo with a computerised digital
  imaging analysis system,''
\newblock {\em Australasian Journal of Dermatology}, vol. 56, no. 4, pp.
  285--289, 2015.

\bibitem{sheth2015pilot}
Vaneeta~M Sheth, Rahul Rithe, Amit~G Pandya, and Anantha Chandrakasan,
\newblock ``A pilot study to determine vitiligo target size using a
  computer-based image analysis program,''
\newblock {\em Journal of the American Academy of Dermatology}, vol. 73, no. 2,
  pp. 342--345, 2015.

\bibitem{roerdink2000watershed}
Jos~BTM Roerdink and Arnold Meijster,
\newblock ``The watershed transform: Definitions, algorithms and
  parallelization strategies,''
\newblock {\em Fundamenta informaticae}, vol. 41, no. 1, 2, pp. 187--228, 2000.

\bibitem{ronneberger2015u}
Olaf Ronneberger, Philipp Fischer, and Thomas Brox,
\newblock ``U-net: Convolutional networks for biomedical image segmentation,''
\newblock in {\em International Conference on Medical image computing and
  computer-assisted intervention}. Springer, 2015, pp. 234--241.

\bibitem{simonyan2014very}
Karen Simonyan and Andrew Zisserman,
\newblock ``Very deep convolutional networks for large-scale image
  recognition,''
\newblock {\em arXiv preprint arXiv:1409.1556}, 2014.

\bibitem{he2016deep}
Kaiming He, Xiangyu Zhang, Shaoqing Ren, and Jian Sun,
\newblock ``Deep residual learning for image recognition,''
\newblock in {\em Proceedings of the IEEE conference on computer vision and
  pattern recognition}, 2016, pp. 770--778.

\bibitem{szegedy2016rethinking}
Christian Szegedy, Vincent Vanhoucke, Sergey Ioffe, Jon Shlens, and Zbigniew
  Wojna,
\newblock ``Rethinking the inception architecture for computer vision,''
\newblock in {\em Proceedings of the IEEE conference on computer vision and
  pattern recognition}, 2016, pp. 2818--2826.

\bibitem{szegedy2017inception}
Christian Szegedy, Sergey Ioffe, Vincent Vanhoucke, and Alexander~A Alemi,
\newblock ``Inception-v4, inception-resnet and the impact of residual
  connections on learning,''
\newblock in {\em Thirty-First AAAI Conference on Artificial Intelligence},
  2017.

\bibitem{hu2018squeeze}
Jie Hu, Li~Shen, and Gang Sun,
\newblock ``Squeeze-and-excitation networks,''
\newblock in {\em Proceedings of the IEEE conference on computer vision and
  pattern recognition}, 2018, pp. 7132--7141.

\bibitem{codella2017deep}
Noel~CF Codella, Q-B Nguyen, Sharath Pankanti, DA~Gutman, Brian Helba,
  AC~Halpern, and John~R Smith,
\newblock ``Deep learning ensembles for melanoma recognition in dermoscopy
  images,''
\newblock {\em IBM Journal of Research and Development}, vol. 61, no. 4/5, pp.
  5--1, 2017.

\bibitem{bergstra2012random}
James Bergstra and Yoshua Bengio,
\newblock ``Random search for hyper-parameter optimization,''
\newblock {\em Journal of Machine Learning Research}, vol. 13, no. Feb, pp.
  281--305, 2012.

\bibitem{snoek2012practical}
Jasper Snoek, Hugo Larochelle, and Ryan~P Adams,
\newblock ``Practical bayesian optimization of machine learning algorithms,''
\newblock in {\em Advances in neural information processing systems}, 2012, pp.
  2951--2959.

\end{thebibliography}

\end{document}